\documentclass[10pt]{article}

\usepackage[a4paper,margin=0.63in]{geometry}
\usepackage[T1]{fontenc}
\usepackage[utf8]{inputenc}
\usepackage{lmodern}
\usepackage{microtype}
\usepackage{booktabs}
\usepackage{array}
\usepackage{enumitem}
\usepackage{url}
\usepackage{hyperref}
\usepackage{ragged2e}
\usepackage{xcolor}
\usepackage{dblfloatfix}
\usepackage{amsmath}
\usepackage[font=footnotesize,labelfont=bf,labelsep=period,skip=3pt]{caption}
\usepackage{titlesec}
\usepackage{tipa}

\hypersetup{
  colorlinks=true,
  linkcolor=black,
  citecolor=black,
  urlcolor=black
}

\setlist[itemize]{leftmargin=*, itemsep=1pt, topsep=2pt}
\setlist[enumerate]{leftmargin=*, itemsep=1pt, topsep=2pt}
\titlespacing*{\section}{0pt}{1.85ex plus 0.35ex minus 0.2ex}{0.72ex}
\titlespacing*{\subsection}{0pt}{1.5ex plus 0.25ex minus 0.2ex}{0.55ex}
\newcommand{\finding}{\mathsf{finding}}
\newcommand{\expect}{\mathcal{E}}
\newcommand{\observation}{\mathcal{O}}
\newcommand{\due}{\mathcal{D}}

\title{\textbf{Intrusion Detection for Agentic Processes: Evidence-Based Runtime Monitoring}}
\author{Arslan Br\"omme\\
\textit{CISSP, CISM, CISA, CAISE}\\
Independent Researcher\\
\texttt{arslan.broemme@aviomatik.de}}
\date{Draft v0.9.1.5 -- 10 September 2026}

\begin{document}
\twocolumn[
\maketitle
\vspace{-1.2em}
\begin{center}
\fbox{%
\begin{minipage}{0.94\textwidth}
\centering
\textit{Preprint / working paper. This version is a work in progress and may be updated. Comments are welcome.}
\end{minipage}}
\end{center}
\vspace{0.45em}
]

\begin{abstract}
Agent deployments increasingly combine language-model inference with retrieval, delegation, tool execution, external-system access, and human approval. Security-relevant deviations can therefore emerge across an evolving process rather than in one isolated input or action. Building on the author's earlier product- and vendor-neutral black-box architecture for agentic processes and the subsequent evidence-claim model, this paper proposes an \emph{Agentic-Process Intrusion Detection System (A-IDS)}, an evidence-aware security interpretation layer for runtime intrusion detection whose monitored object is the agentic process itself. A-IDS compares evidence-supported observations with an explicitly governed and versioned expectation baseline for workflow state, authorization, communication, and mandatory events. Its conceptual contribution combines dynamically due governed expectations, visibility separated from three-valued matching, explicit unresolved observation states, and bounded findings that separate evidentiary status from operational impact. The model further identifies the monitoring plane itself as an attack surface when adversarial content reaches semantic evidence producers through otherwise legitimate observation paths. Some observations may be produced outside the operational agent's self-report path, but the model does not assume complete observability or universally trustworthy capture. Prompt injection is treated not only as an input-security problem but also as a possible origin of later process deviations and cross-agent influence paths. A-IDS does not infer malicious intent from anomalous behavior, does not treat an unobserved event as proof of non-occurrence, and does not claim a new anomaly detector, temporal logic, or provenance model. The contribution is conceptual: it does not validate a particular implementation, demonstrate empirical detection performance, establish causal attribution, or provide an enforcement mechanism.
\end{abstract}
\vspace{0.55em}

\section{Introduction}

An \emph{agentic process} is a bounded task-oriented sequence or graph of observable interactions among AI agents, humans, tools, external systems, and governance or evidence components. Such processes can retrieve external information, delegate work, invoke tools, update artifacts, and operate under changing policy and approval states. A security-relevant deviation may therefore emerge across several otherwise plausible events, for example untrusted content exposure, changed inter-agent communication, an omitted approval observation, and a privileged tool invocation.

Traditional intrusion detection and prevention systems monitor hosts, networks, applications, or communication behavior for possible incidents~\cite{nist80094}. Agentic processes add semantic workflow state, delegated authority, tool use, approval obligations, and multi-agent communication. The relevant question is therefore not only whether a host or flow is anomalous, but whether the evolving agentic process is supported by evidence consistent with its expected secure state.

The author's earlier black-box architecture proposed a product- and vendor-neutral evidence pipeline for agentic processes based on capture, canonicalization, hashing, external anchoring, certification, and verification~\cite{broemme2026blackbox}. A subsequent evidence-model paper defined bounded evidence claims and emphasized that integrity, provenance, authorization, completeness, and semantic validity are distinct properties~\cite{broemme2026evidence}. It also introduced an evidence supervisor, independently defined expected events, and a mandatory-event coverage indicator. The present paper extends that line from evidence preservation and interpretation toward runtime security detection.

The research question is: \emph{How can security-relevant deviations in an agentic process be detected at runtime from evidence-supported observations without treating anomalous behavior as proof of malicious intent?}

The contribution is not the general idea of agent monitoring. A-IDS\footnote{Pronunciation (IPA): \textipa{/eI | aI di: Es/}.} is not proposed as a new anomaly detector, temporal logic, provenance model, or policy-enforcement mechanism. It defines an evidence-aware security interpretation layer that (1) evaluates dynamically due governed expectations, (2) separates visibility from event matching, (3) represents unresolved evidence explicitly, (4) emits bounded findings that separate evidentiary status from operational impact, and (5) identifies adversarial observation content as a distinct monitoring-plane attack surface and derives isolation and trust-boundary requirements for semantic evidence producers. In this paper, intrusion detection follows the operational framing of Intrusion Detection and Prevention Systems (IDPS), i.e., the identification of possible security incidents or security-relevant deviations~\cite{nist80094}. A policy deviation is not automatically an intrusion, and a finding does not establish attack, compromise, intent, causality, or legal non-compliance.

\section{Background and Related Work}

Classical IDPS distinguishes monitoring locations and detection methods, including signature, anomaly, and stateful protocol analysis. A-IDS is closest to specification- and state-aware monitoring because it evaluates observed process events against expected conditions. Specification-based intrusion detection has a longer history in classical security. Sekar et al. combined state-machine specifications of network protocols with anomaly detection to constrain expected behavior while retaining sensitivity to novel attacks~\cite{sekar2002}. A-IDS draws only the general monitoring analogy. Its governed expectations, bounded evidence semantics, and visibility model concern agentic processes rather than network-protocol state. Signature, anomaly, graph, and content detectors can contribute upstream evidence without becoming the novel core of A-IDS.

The comparison of event logs with expected process behavior is established in process mining and conformance checking~\cite{rozinat2008,vanderaalst2011}. Runtime verification similarly evaluates partial executions against formal properties, including temporal constraints and inconclusive states~\cite{bauer2011}. Security log management and provenance standards provide complementary foundations for record lifecycle, source relationships, and later reconstruction~\cite{nist80092,w3cprov}. A-IDS does not propose a new temporal logic, process language, or provenance standard. Its narrower focus is the security interpretation of conformance deviations when observation visibility and evidentiary support remain bounded and the expectation baseline is itself governed.

Recent agent-security work addresses adjacent problems. SentinelAgent represents multi-agent interactions as execution graphs and detects anomalies at node, edge, and path levels~\cite{sentinelagent}. VIGIL evaluates agent execution traces against behavioral specifications with temporal, argument, and value-flow constraints~\cite{vigil}. C-Trace expresses selected compliance requirements as predicates over runtime traces and enforces selected regulatory constraints~\cite{ctrace}. Compliance monitoring is adjacent but distinct. A runtime deviation from a machine-testable control can support later compliance assessment, but it does not by itself establish legal non-compliance. The same expectation-and-evidence structure may therefore be reusable for compliance deviations without changing the security scope of A-IDS. An OpenAI industry disclosure describes low-latency monitoring of internal coding-agent trajectories for suspicious or policy-inconsistent behavior~\cite{openaiMonitor}. AttriGuard evaluates whether tool calls are causally supported by user intent or driven by untrusted observations~\cite{attriguard}. NeuroTaint reconstructs untrusted-source to privileged-sink information flow as an offline auditor, making it relevant as a post-hoc or nearline evidence producer rather than a synchronous pre-action detector~\cite{neurotaint}.

MITRE ATLAS and OWASP's Agentic Applications Top 10 provide threat taxonomies for AI and agentic systems, while the NIST AI RMF provides a broader governance context for managing AI risk~\cite{mitreatlas,owaspagentic,nistairmf}. NIST incident-response guidance helps distinguish a runtime finding from later incident analysis and response~\cite{nist80061r3}. These frameworks motivate threat coverage and governance, but do not by themselves define the runtime evidence semantics proposed here. AgentDojo and InjecAgent provide established environments for prompt-injection evaluation and are relevant candidates for that part of later empirical validation~\cite{agentdojo,injecagent}. They do not by themselves evaluate the broader A-IDS dimensions of process deviation, authorization, visibility, or evidence-path monitoring.

\section{A-IDS Model}

\subsection{Observations, Trust, and Threat Model}

The monitored process is represented through observable events rather than inferred hidden reasoning. We distinguish an \emph{event} in the monitored world, a \emph{record} that technically captures it, an \emph{observation} that normalizes or validates one or more records, an \emph{evidence claim} that states a bounded property of those records or observations, and a \emph{security finding} that compares such claims with governed expectations. Let \(\observation\) denote the observation space. A concrete observation \(o\in\observation\) should bind at least event type, actor identity, relevant object or counterparty, local time or sequence reference, policy or workflow version, capture source, and an evidence reference or content commitment. Typical events include agent-agent messages, tool requests and results, policy checks, approvals, delegation, credential use, external access, artifact changes, and evidence-path events. A minimal observation can be represented as

\[
\begin{aligned}
o=(&\text{type},\text{actor},\text{object},\text{time/sequence},\\
&\text{source},\text{policy version},\text{evidence reference}).
\end{aligned}
\]

A source should not be treated as independent merely because it is distinct from another source. Such a claim requires additional criteria concerning source separation, control, authentication, integrity protection, and shared failure modes. An agent self-report differs from a gateway-observed event, an external-system receipt, or a source-separated process observer. A cryptographically committed record provides integrity evidence, not semantic truth or completeness. Cross-source corroboration can strengthen support when the sources do not share the same failure mode. Whether a source is operationally separate, externally controlled, authenticated, committed, or corroborated is therefore recorded as a distinct property. Here, \emph{source health} denotes whether an observation source was operational and sufficiently reliable as a capture path during the relevant interval, including the absence of known outages, capture gaps, or unresolved operational faults. Source health does not establish the semantic truth of captured data.

The threat model includes a compromised or misdirected operational agent, adversarial external content, a malicious peer agent, a compromised tool or endpoint, privileged-user misuse, and manipulation of some capture or monitoring components. A-IDS therefore considers two related but distinct attack surfaces: deviations within the monitored agentic process and attacks against the monitoring plane through adversarial observation content. An adversary may attempt to spoof plausible observations, artificially indicate sufficient coverage, alter sequence information, or adapt behavior to evade detection rules. The model therefore assumes that the attacker does not simultaneously control every relevant observation source and the authoritative expectation-definition path. If the monitored process, all observation paths, and the governing expectation baseline are controlled by the same adversary, A-IDS cannot provide meaningful detection assurance.

\subsection{Governed Expectations and Matching}

For A-IDS, a security policy \(P_v\) is treated as a versioned set of governed security expectations and authorization relations, together with their scope, applicability conditions, evidence requirements, exception rules, and governance metadata. A-IDS does not prescribe a specific security-policy standard or language. In practice, governed expectations may be derived from organizational security policies, ISMS requirements such as ISO/IEC 27001 and BSI IT-Grundschutz, control catalogs such as NIST SP 800-53, or prioritized security baselines such as the CIS Controls, provided that relevant requirements can be translated into versioned, machine-testable expectations~\cite{iso27001,nist80053,bsi2001,ciscontrols}.

Let \(\expect\) denote the expectation space. A concrete expectation \(e\in\expect\) is a versioned rule describing an observable obligation or allowed relation. We model it compactly as

\[
\begin{aligned}
e=(&\text{type},\text{actor},\text{object},\text{trigger},\text{precedence},\\
&\text{deadline},\text{cardinality},\text{source requirement},\\
&\text{policy version}).
\end{aligned}
\]

This tuple is an interface, not a new workflow language. A deployment may instantiate it using state machines, declarative constraints, temporal logic, policy tables, or tool-gateway rules. Cardinality can express, for example, exactly one authorized approval or any one of several approved sources. A source requirement can demand a gateway observation, external receipt, or corroboration across specified source classes. Expectation definitions should themselves be governed evidence objects with an issuer or owner, version, effective interval, authorization record, integrity commitment, and supersession relation. A runtime finding must bind to the authorized rule version effective for the relevant process window. For processes already in execution, policy should specify the applicable version or record an authorized transition event so that earlier actions are not evaluated retroactively under a later baseline.

Matching is three-valued. For policy version \(P_v\), let
\[
\operatorname{match}_{P_v}(o,e)\in\{\text{true},\text{false},\text{unknown}\}.
\]
A result is true only when the policy-defined type, actor, object, identity, temporal, cardinality, and source constraints are satisfied. False means that sufficiently resolved attributes violate a matching constraint. Unknown is required when identity, timing, source quality, or event granularity is unresolved. Matching is structural by default. A semantic classifier does not create a true match by itself. Its model version, input, output, calibration status where available, and error assumptions are retained as a bounded evidence claim, and the governing policy decides how that claim may contribute to matching. Mandatory obligations are matched one-to-one by default. Alternatives, repeated events, compensation, record aggregation, or one observation satisfying several expectations require explicit versioned normalization or matching rules. Duplicate records are deduplicated by governed identifiers or commitments. Late arrivals may reclassify a state but do not erase the earlier monitoring history. Clock skew is handled by a policy-defined tolerance window, with event time distinguished from processing time.

Visibility is evaluated separately from matching. Let \(V_t(e)\in\{\text{Sufficient},\text{Insufficient},\text{Unknown}\}\) state whether the required observation paths for expectation \(e\) were sufficiently available by runtime point \(t\). Visibility is Sufficient only when every source class required by the expectation was active and in scope for the relevant interval, the required capture path was enabled for the event class in question, no documented capture gap or known event-class omission affects that interval, source health is not Unknown, and required identity and time bindings are resolved within the governing tolerance. Source health is therefore necessary but not sufficient for Sufficient visibility. Identity uncertainty or clock skew outside the governing tolerance prevents a Sufficient classification. It is Insufficient when a known gap, failed required source, or known event-class omission prevents that conclusion, and Unknown when source health, scope, or binding cannot be established. A due expectation is \textbf{Observed} when a validated true match exists, \textbf{Unobserved} when no true match exists and \(V_t(e)=\text{Sufficient}\), \textbf{Violated} when resolved observations establish a policy-defined contrary condition, and \textbf{Unknown} when visibility or matching remains unresolved. Conflicting observations from different sources should not, by themselves, be classified as a violation. Instead, they should generate an Evidence Conflict finding. A governed source-resolution rule may still support an Observed state for the obligation, while an unresolved conflict normally leaves the affected expectation Unknown, as summarized in Table~\ref{tab:matchvisibility}.

\par\smallskip
\begin{center}
\footnotesize
\renewcommand{\arraystretch}{1.0}
\begin{tabular}{>{\RaggedRight\arraybackslash}p{0.35\linewidth} >{\RaggedRight\arraybackslash}p{0.21\linewidth} >{\RaggedRight\arraybackslash}p{0.26\linewidth}}
\toprule
\textbf{Match / evidence} & \textbf{Visibility} & \textbf{State} \\
\midrule
true match, no unresolved required conflict & Sufficient & Observed \\
true match under source-resolution rule + source conflict & Sufficient & Observed; separate Evidence Conflict \\
no true match & Sufficient & Unobserved \\
no true match & Insufficient or\newline Unknown & Unknown \\
resolved contrary condition & Sufficient & Violated \\
unknown match & any & Unknown \\
unresolved source conflict & any & Unknown; separate Evidence Conflict \\
\bottomrule
\end{tabular}
\captionof{table}{Match and visibility state mapping.}
\label{tab:matchvisibility}
\end{center}
\smallskip

Let \(\due_t\) be the set of mandatory expectations whose trigger has occurred and whose ordering or deadline makes them testable by runtime point \(t\). Let \(s_t(e)\) denote the observation state of expectation \(e\) at runtime point \(t\), with \(s_t(e)\in\{\text{Observed},\text{Unobserved},\text{Violated},\text{Unknown}\}\). Let \(M_t=|\{e\in\due_t:s_t(e)=\text{Observed}\}|\). The due-event match ratio is

\[
C_t=\frac{M_t}{|\due_t|}, \qquad |\due_t|>0.
\]

When \(|\due_t|=0\), \(C_t\) is not applicable rather than zero or one. Unknown and Violated expectations remain in the denominator but not the numerator and should be reported separately from Unobserved expectations. The ratio is therefore a narrow matching signal, not a completeness or correctness proof. Due-state transitions may be event-driven, deadline-driven, or policy-exception-driven. Parallel branches remain separate until an explicit join. Long-lived obligations retain their due state across monitoring windows. A compensating action satisfies only a separately defined compensation expectation and does not erase the earlier deviation. Policy changes during execution follow the version applicable to the process or an evidenced transition event. The following five design properties are intended requirements for an implementation and are not proven implementation properties in this conceptual paper.
\begin{enumerate}[label=(\roman*),leftmargin=1.6em,itemsep=0pt,topsep=1pt,parsep=0pt]
\item no future-event penalty
\item no inference of non-occurrence from Unobserved
\item baseline binding to the version effective in the process window
\item preservation of source conflicts
\item monotonic evidence history, in which later records may reclassify or supersede findings but do not erase earlier evidence
\end{enumerate}

\subsection{Detection Taxonomy}

Table~\ref{tab:detection} groups the initial detection classes by the type of deviation rather than treating them as one homogeneous ontology.

\begin{table*}[!t]
\centering
\footnotesize
\renewcommand{\arraystretch}{1.04}
\begin{tabular}{p{0.16\textwidth} >{\RaggedRight\arraybackslash}p{0.18\textwidth} p{0.27\textwidth} p{0.28\textwidth}}
\toprule
\textbf{Category} & \textbf{Detection class} & \textbf{Observed condition} & \textbf{Bounded interpretation} \\
\midrule
Process & Unobserved mandatory event & A due obligation is Unobserved. & Supports an observation- or process-gap finding. Does not prove non-occurrence. \\
Process & Sequence deviation & Events violate required ordering, state, or preconditions. & Supports workflow deviation, not malicious intent or physical causality. \\
Authorization & Unauthorized interaction & Agent contacts an agent, tool, endpoint, or data source outside an allowed relation. & Supports an interaction-policy finding if identity and policy binding are reliable. \\
Escalation & Behavioral escalation & After denial or restriction, repeated or redirected actions target the same restricted resource or operation. & Supports circumvention if explicit retry or compensation rules do not permit the path. \\
Evidence & Evidence conflict & Relevant observations are mutually inconsistent. & Supports a consistency finding. It does not identify the correct source. \\
Evidence & Evidence-path anomaly & Capture, sequencing, supervision, or evidence-control obligations show gaps or bypass indicators. & Supports a finding against observability itself and weakens downstream assurance. \\
Influence & Source-associated deviation & A deviation follows untrusted exposure and a traceable source-to-action path exists. & Supports a bounded influence hypothesis. Association alone does not prove causal control. \\
\bottomrule
\end{tabular}
\caption{Initial A-IDS deviation classes and bounded interpretations.}
\label{tab:detection}
\end{table*}

The classes may overlap and form a multidimensional taxonomy comprising deviation domain, observation or evidence state, and operational impact, which remain distinct. A privileged tool invocation without observed required approval can be both an Unobserved mandatory event and a sequence deviation. Unauthorized interaction is evaluated against an explicit authority relation covering actor, delegated principal, permitted tool or counterparty, data domain, temporal scope, and where relevant purpose. Technical success does not prove authorization. This includes confused-deputy patterns in which an otherwise authorized component exercises authority for an unauthorized or mismatched principal~\cite{hardy1988}. Behavioral escalation is structural by default, such as repeated access to the same restricted resource or operation. Semantic equivalence requires a governed classifier whose output remains a bounded evidence claim. If a deviation follows adversarial content exposure, it may also support a source-associated finding. These components should remain separately represented and linked. They should not be combined into an unsupported conclusion that an agent was compromised.

\subsection{Auditable Finding Structure}

A finding should preserve both the evidentiary basis and the operational importance of the deviation. We model an A-IDS finding as
\begingroup
\setlength{\abovedisplayskip}{2pt}
\setlength{\belowdisplayskip}{2pt}
\fontsize{8.6pt}{9.2pt}\selectfont
\[
\begin{aligned}
\finding=(&\text{window},\text{expectation},\text{observation},\\
&\text{deviation},\text{origin hypothesis (optional)},\\
&\text{support},\text{evidence status},\text{impact},\\
&\text{assumptions},\text{limitation}).
\end{aligned}
\]
\endgroup

\emph{Observation} identifies the concrete records used by the finding. \emph{Evidence status} summarizes their observation state, visibility, source separation, validation, and consistency. \emph{Support} binds record, observation, source, matching, visibility, classifier, and trust references used to justify the finding. \emph{Impact} expresses potential operational consequence or escalation priority. These dimensions must remain separate. Strong evidence for a minor deviation is not equivalent to weak evidence for a potentially destructive action. The optional origin hypothesis is absent unless minimum source-to-action support exists. The structure is deliberately not a probability model, and evidence status must not be interpreted as the probability that an agent is compromised. A practical finding record should also carry a finding ID, process or correlation ID, monitor version, emission time, policy version, and lifecycle status. A lifecycle such as provisional, corroborated, resolved, superseded, or withdrawn may update interpretation without deleting the original record. Overlapping findings may share an actor, process window, evidence commitment, or correlation identifier while retaining separate bounded interpretations. Incident-level aggregation and composite severity are left to later incident-analysis work. Until then, downstream queues may sort individual findings by impact while displaying evidence status separately. This is prioritization, not incident-level severity. Structured findings can be forwarded to SIEM or SOAR pipelines without granting the detector unrestricted enforcement authority.

\section{A-IDS Architecture and Pluggable Evidence Producers}

The novel core of A-IDS is the expectation-and-evidence finding layer, not a claim to implement every classical detector. Its specification- and state-aware logic evaluates governed expectations against observations. Signature, anomaly, content, graph, and information-flow detectors are optional upstream \emph{evidence producers}. Their outputs can identify known injection patterns, unusual communication, new interaction paths, or possible influence, but they enter A-IDS as bounded claims with their own assumptions. Evidence-path monitoring separately checks whether observation infrastructure shows gaps, conflicts, or bypass indicators.

A practical deployment can separate five functions. \textbf{Sensors} observe agent, human, tool, gateway, and external-system events. \textbf{Evidence processing} authenticates and normalizes observations and records their source properties and commitments. A governed \textbf{expectation registry} supplies the authorized baseline and its effective versions. A \textbf{matching and correlation layer} evaluates due expectations, observation states, evidence conflicts, and optional detector claims. Finally, a \textbf{finding function} emits the bounded record defined above. The resulting finding can be forwarded to a SOC, SIEM, SOAR workflow, human reviewer, audit function, or a separately authorized prevention layer.

\begin{center}
\footnotesize
\fbox{\begin{minipage}{0.93\linewidth}
\centering
Sensors $\rightarrow$ evidence processing $\rightarrow$ matching / correlation\\
Governed expectation registry $\rightarrow$ matching / correlation\\
$\rightarrow$ bounded A-IDS finding $\rightarrow$ SOC / SIEM / audit / later prevention
\end{minipage}}
\end{center}

\subsection{Deployment and Monitoring-Plane Isolation}

A-IDS is intended to complement, not replace, existing security-monitoring infrastructure. Alerts and telemetry from IDS, EDR, NDR, API or tool gateways, and other controls may enter A-IDS as observations or bounded evidence-producer outputs, where they can be correlated with process state, authorization relations, governed expectations, and other sources. Resulting findings can be forwarded to SIEM, SOC, or SOAR workflows, enabling an agent-process-aware detection layer without replacing established monitoring or incident-handling systems.

A practical deployment should separate the monitored agent process from the A-IDS monitoring plane. The monitored process should have no authority to modify the monitor's policy baseline, configuration, credentials, evidence store, or execution state. Observation interfaces can therefore be logically one-way or read-only: defined telemetry and artifacts cross the boundary, while monitored actors receive no control over the monitor. This reduces direct control-plane manipulation without making the monitoring system immune to adversarial input.

This separation is also a trust boundary. Components that observe actions should not rewrite the expectations against which those actions are judged, and a monitor issuing findings should not modify its own detection policy in response to observed events. Expectation changes should follow a governed lifecycle such as propose, authorize, activate, supersede, and retain, and should create evidence records. For higher assurance, registry history can be committed and externally anchored through the black-box pipeline described in prior work~\cite{broemme2026blackbox}. Registry integrity and authorization remain distinct: a malicious policy can be correctly recorded and anchored. High-impact changes may additionally require separate keys, multi-party authorization, or independent review, while emergency changes should not erase earlier deviations.

\subsection{Monitoring-Plane Attacks and Layered Defense}

Content carried through a permitted observation path may itself target an agentic monitoring component. Indirect prompt injection has shown that instructions embedded in data processed by an LLM-integrated application can influence its behavior~\cite{greshake2023indirectprompt}. In A-IDS, an external document may reach a semantic analyzer, while a compromised monitored agent may more deliberately create or modify a legitimate artifact so that it contains instructions intended for the monitor. The producing action may be fully authorized, correctly sequenced, and process-conformant. \emph{Process conformance does not imply that content produced by the conformant process is safe for the monitoring system to interpret.} A monitored actor may therefore attack the monitoring system without first producing a detectable workflow or authorization deviation.

One-way communication does not remove this observation-path attack surface. \emph{One-way observation can protect the monitoring control plane from direct modification, but it does not prevent adversarial content from reaching semantic analysis through the permitted observation path.} Content originating from the monitored process should therefore be treated as potentially adversarial evidence input even when produced by an authorized, process-conformant action.

For higher-assurance deployments, semantic interpretation should be separated from privileged monitoring functions. Hardened components can first capture and normalize raw observations. An LLM-based analyzer can then operate as an isolated evidence producer whose output remains a bounded claim, with no authority over the expectation registry, monitoring configuration, credentials, retained evidence, finding history, or enforcement. Governed matching and finding generation can remain separate from potentially adversarial content interpretation. This limits, but does not eliminate, prompt-injection risk.

Isolation bounds authority, not trust. A manipulated semantic claim may still influence matching through a governance-authorized path. Rules that admit classifier claims into matching should therefore preserve their source and trust properties. A classifier processing content from the monitored process should not be treated as source-separated confirmation of that same process. A-IDS must therefore bound both the authority of semantic analyzers and the evidentiary weight assigned to their outputs.

\subsection{Distributed Correlation and Source Separation}

Distributed correlation is particularly useful when no single sensor is authoritative. Consider a technical conflict: an agent self-report states that a privileged tool call failed, a gateway records successful dispatch, and the external system records receipt. If the governing expectation accepts gateway plus external receipt as the required source classes, the tool invocation can be Observed while the contradictory self-report creates a separate Evidence Conflict finding. If no governed source-resolution rule exists for the disputed condition, the affected expectation remains Unknown unless a governed resolution rule supports classification as Observed or Violated. Several records can also share one infrastructure and one failure mode, so corroboration is strongest when observations cross meaningful trust boundaries and do not depend on the same potentially compromised source.

\section{Prompt Injection and Influence Paths}

Prompt injection matters to A-IDS both as a detectable input condition and as a possible origin of later deviations, consistent with its treatment as an information-security risk in the NIST Generative AI Profile~\cite{nistgai}. Consider the expected sequence

\begin{center}
\small
\fbox{\begin{minipage}{0.93\linewidth}
External content $\rightarrow$ policy check $\rightarrow$ risk assessment $\rightarrow$ human approval $\rightarrow$ privileged tool call $\rightarrow$ verification
\end{minipage}}
\end{center}

Suppose an external document contains a suspected approval-bypass instruction. The external-content exposure is Observed, as is Agent A's message to Agent B. A privileged tool call is later Observed. If the required approval has no matching record and the mandated approval-capture paths have Sufficient visibility, that expectation is Unobserved. If the observed tool call also violates a required ordering rule, the sequence expectation is Violated. A-IDS can bind these states to the governing baseline and may add a source-associated deviation only when the origin criteria below are met. The finding must still state that the evidence does not prove that the document caused the action, that no off-system approval existed, or that either agent had malicious intent.

An origin hypothesis requires more than temporal proximity. At minimum, A-IDS should retain an identifiable untrusted source, evidence of exposure or propagation toward the affected actor, and a temporally compatible path to the deviation. Shared or semantically related content can strengthen support only when tied to a traceable propagation or information-flow claim. Semantic similarity alone is insufficient. AttriGuard can contribute causal-attribution claims, while NeuroTaint contributes offline source-to-sink provenance claims~\cite{attriguard,neurotaint}. Their outputs remain bounded evidence with explicit assumptions and do not establish semantic truth. \emph{Source-associated} denotes an evidentiary relationship, not an established causal relationship, attack origin, or proof of agent control.

\section{Detection, Auditability, and Evaluation}

Runtime detection, incident analysis, and audit use related evidence but answer different questions. A-IDS asks whether the current process exhibits a security-relevant deviation. Incident analysis reconstructs source, propagation, decisions, actions, and consequences. Audit evaluates retained evidence against policy or control requirements. A security finding should not become an incident or compliance conclusion merely because the same evidence later supports response or audit~\cite{nist80061r3}.

Runtime findings should themselves enter the evidence stream. A finding record should bind monitor identity and version, authorized expectation version, process window, input evidence commitments, output finding, evidence status, impact, and escalation. This creates second-order evidence, but also a regress risk. A compromised monitor can issue a well-formed false finding. Higher-assurance deployments should therefore segregate monitoring from operational execution, govern monitor configuration and keys independently, record monitor changes, correlate multiple observation sources where feasible, and apply periodic human or external sampling.

The paper is a conceptual design and does not provide empirical validation. A minimal evaluation should inject approval bypass, unauthorized interaction, prompt-injection propagation, and evidence-path manipulation into realistic agent workflows. Ground truth should be defined over instrumented traces and injected control conditions, rather than over unobservable intent or whether an agent was in fact compromised. Core questions are whether due-event and sequence deviations are detected under complete visibility, whether explicit visibility reduces unsupported claims under partial capture, whether evidence-path monitoring exposes degraded assurance, and what runtime and storage overhead result. Measures should include precision, recall, false-positive rate, detection latency, observation loss, robustness under conflicting evidence, and resistance to monitor-targeted evasion. Baselines can include a single-call filter, pure anomaly detection, runtime specification monitoring, A-IDS without evidence-path monitoring, and the full model. Ablations should remove visibility, source properties, governed versioning, evidence-path monitoring, or three-valued matching. Evidence-manipulation ground truth should originate outside the manipulated pipeline, for example from a signed scenario manifest or external test harness. AgentDojo and InjecAgent provide prompt-injection starting points~\cite{agentdojo,injecagent}.

\subsection{Limitations}
False positives (FPs) can arise from incomplete specifications, delayed events, delegation ambiguity, or legitimate adaptation. False negatives (FNs) can arise from off-system actions, policy gaps, observer compromise, semantic transformation, collusion, or monitor-targeted evasion. Table~\ref{tab:failuremodes} summarizes representative causes. They primarily affect evidence status and assurance, while impact remains separate. Runtime observability should be balanced against data minimization, access control, retention, and content commitments where feasible~\cite{nist80092}. Event time and processing time must remain distinct, with explicit handling for late, duplicate, replayed, or reordered records. Cost depends on active expectations, concurrent processes, correlation fan-out, and optional semantic classifiers. Cross-organization deployments require explicit agreement on expectation authority, evidence disclosure, and trust boundaries.

\begin{table}[!t]
\centering
\fontsize{6.2pt}{6.6pt}\selectfont
\renewcommand{\arraystretch}{0.96}
\begin{tabular}{p{0.28\linewidth} p{0.61\linewidth}}
\toprule
\textbf{Mode} & \textbf{A-IDS implication} \\
\midrule
Specification FP & Valid finding against an incomplete baseline. \\
Timing FP & Late evidence may reclassify Unobserved or Unknown. \\
Visibility FN & Side-channel activity may bypass all observers. \\
Policy-gap FN & Harmful activity may be outside governed expectations. \\
Spoofing / evasion & Plausible records or rule-aware behavior may evade detection. \\
Collusion / semantic change & Cross-agent or paraphrased influence may defeat exact signatures. \\
\bottomrule
\end{tabular}
\caption{Representative false-positive, false-negative, and evasion modes.}
\label{tab:failuremodes}
\end{table}

\section{Conclusion and Future Work}

A-IDS extends intrusion detection to the agentic process as a monitored object. Its distinctive unit is not merely an anomaly score or policy violation, but a bounded, auditable security finding that links an authorized expectation to evidence-supported observations, deviation type, evidence status, impact, assumptions, and limitations. The expectation baseline is itself governed and versioned, while observation uncertainty is represented explicitly through visibility, three-valued matching, and the states Observed, Unobserved, Violated, or Unknown.

The central principle is that detection remains bounded by evidence. An unobserved approval is not proof that approval never occurred. A source-associated deviation is not proof of causal prompt-injection control. A severe action with weak evidence is not equivalent to a minor deviation with strong evidence. These distinctions allow runtime escalation without making claims that exceed the available evidence.

Future work should implement and evaluate the model under partial observability, adversarial attempts to evade or manipulate monitoring, policy drift, parallel workflows, and conflicting evidence. A direct continuation of this work will examine an \emph{Agentic-Process Intrusion Prevention System (A-IPS)}, in which evidence-based A-IDS findings drive separately authorized and proportionate preventive or containment actions, from reinforcement and reauthorization through restriction, isolation, credential revocation, quarantine, or termination. A further line will examine runtime detection and treatment of compliance deviations while keeping technical deviation findings distinct from legal conclusions of non-compliance. Evidence-based incident analysis can then reconstruct source, propagation, decision, action, and consequence paths after an alert.

\vspace{-0.35em}
\begingroup
\setlength{\fboxsep}{2.0pt}
\noindent\hspace*{0.60em}\fbox{%
\begin{minipage}{0.91\linewidth}
\fontsize{5.8pt}{6.2pt}\selectfont
\textbf{Declaration on the Use of AI Tools.} AI language models, in particular OpenAI GPT-5.6 and Anthropic Claude Sonnet 5, were used as tools during the preparation of this paper for language drafting, critical review, source checking, discussion of examples, and LaTeX/PDF artifact generation. The author remains solely responsible for the content, conceptual decisions, source selection, and final version.
\end{minipage}}
\endgroup

\end{document}